\documentclass[a4paper,10pt]{article}

\usepackage{graphicx}

\title{Compressibility of a two-dimensional extended Hubbard model}
\author{
E. J. Calegari and S. G. Magalh\~aes\footnote{ggarcia@ccne.ufsm.br} \\ \\
{\it Laborat\'orio de Mec\^anica Estat\'{\i}stica e Teoria da Mat\'eria Condensada}\\
{\it Universidade Federal de Santa Maria, 97105-900 Santa Maria, RS, Brazil}\\  \\
A. A. Gomes \\ \\
{\it Centro Brasileiro de Pesquisas F\'{\i}sicas-CBPF}\\
{\it  Rua Xavier Sigaud 150, 22290-180, Rio de Janeiro, RJ, Brazil}}

\begin{document}

\maketitle

\begin{abstract}
The compressibility of an extended $d-p$ Hubbard model is investigated by the Roth's 
two-pole approximation. Using the factorization procedure proposed by Beenen and Edwards, 
superconductivity with singlet $d_{x^2-y^2}$-wave pairing is also considered.  Within this 
framework, the effects of $d-p$ hybridization and Coulomb interaction $U$ on the 
compressibility are studied carefully. It has been found that the compressibility diverges 
and then it becomes negative near the half-filling. Within Roth's method, it has been verified 
that an important contribution for the negative compressibility comes from the spin-correlation 
term $\langle S_j^+S_i^-\rangle$ present in Roth's band shift. This correlation function plays an 
important role due to its high doping dependence. Also, its effects in the band shift and 
consequently in the compressibility are pronounced near the half-filling. 
The numerical results show that the hybridization acts in the sense of suppressing the negative 
compressibility near half-filling. Finally, the possibility of a connection between the negative 
compressibility and the phase separation is also discussed.   

\end{abstract}
\vspace{3.0cm}

The non-Fermi liquid behavior observed in the normal state of the underdoped
regime in cuprate systems has been subject of plenty study in recent years \cite{ref00}. 
The existence of pseudogap, stripes and phase separation \cite{ref00,ref01} in the normal 
state, becomes appreciably hard the 
experimental and theoretical understanding of these systems. As the cuprates are strongly 
correlated electron systems, it is believed that the one-band Hubbard model can capture their main 
physical properties\cite{ref01}. However, although this model gives interesting results, 
probably, a model which takes into account the hybridization between the $d-$ and $p-$orbitals
can be more adequate to treat these systems \cite{ref4}. Recently, the present authors have applied 
the Roth's two-pole approximation \cite{ref5} to study the effects of $d-p$ hybridization on 
the superconducting properties of an extended $d-p$ Hubbard model \cite{ref4,ref2,ref3}. 
The authors followed the factorization procedure proposed by Beneen and Edwards \cite{ref52} 
to consider $d-$wave superconductivity. 
However, near the half-filling, the obtained results 
show that the compressibility ($\kappa=\frac{dn_T}{d\mu}$) diverges in a critical value 
of total occupation number $n_T$ (where $n_T=n_{\sigma}^d+n_{-\sigma}^d$), and then, it 
becomes negative. Nevertheless, the divergence may be associated with phase separation which 
occurs in the underdoped regime of the cuprate systems \cite{ref01}. In this work, the nature of 
the divergence and the 
negative compressibility has been carefully investigated in the framework of the Roth's 
approximation \cite{ref5}. The model used \cite{ref4} is given by:  
\begin{eqnarray}
H&=&\sum_{i,\sigma }(\varepsilon _{d}-\mu)d_{i\sigma }^{\dag}d_{i\sigma
}+\sum_{i,j,\sigma }t_{ij}^{d}d_{i\sigma }^{\dag}d_{j\sigma }+
U\sum_{i}n_{i\uparrow}^{d}n_{i\downarrow}^{d}\nonumber\\
 & &+\sum_{i,\sigma }(\varepsilon _{p}-\mu)p_{i\sigma }^{\dag}p_{i\sigma
}+\sum_{i,j,\sigma }t_{ij}^{p}p_{i\sigma }^{\dag}p_{j\sigma }\nonumber\\
 & &+\sum_{i,j,\sigma }t_{ij}^{pd}\left( d_{i\sigma
}^{\dag}p_{j\sigma +}p_{i\sigma }^{\dag}d_{j\sigma }\right)
\label{eq1}
\end{eqnarray}
where $\mu$ is the chemical potential.
The Green's functions necessary to treat the problem were obtained following the 
Roth's standart procedure \cite{ref5}, and they are found in reference \cite{ref4}. 
The Roth's band shift 
$W_{{\bf k}\sigma}=W_{{\bf k}\sigma}^d+W_{\sigma}^{pd}$ is given in terms of correlation
functions which can be evaluated from the obtained Green's functions (see Refs. \cite{ref4,ref3}). 
The $W_{{\bf k}\sigma}^d$ is given as:
\begin{equation}
n_{-\sigma}^d(1-n_{-\sigma}^d)W_{{\bf k}\sigma }^d=h_{1\sigma}
+\sum_{j\neq 0}t_{0j}^{d}e^{{i{\bf k}}\cdot{{\bf R}_j}}\overline{h}_{j\sigma}
\label{eq13}
\end{equation}
where $\overline{h}_{j\sigma}=h_{2j\sigma}+h_{3j\sigma}$ and  $j$ stand for the nearest 
neighbors of the $i$ site. The term $h_{1\sigma}$ is 
$n_T$-dependent but it is $\bf k$-independent.  The second term in the right side of 
Eq. (\ref{eq13}) is both $n_T$ and $\bf{k}$-dependent. As discussed in Ref. \cite{ref4}, 
the term $h_{3j\sigma}$, which is directly associated with the superconductivity,
is quite small and therefore, it can be disregarded. Finally, 
$h_{2j\sigma}=F_{j\sigma}^{(1)}+F_{j\sigma}^{(2)}+F_{3j\sigma}^{(3)}$ with, 
\begin{eqnarray}
F_{j\sigma}^{(1)}&=&-\frac{\alpha_{j\sigma}n_{0j\sigma}^d + \beta_{j\sigma}m_{j\sigma}}
{1-\beta_{\sigma}\beta_{-\sigma}}\label{eq14}\\
F_{j\sigma}^{(2)}&=&\langle S_j^+S_0^-\rangle=-\frac{\alpha_{j\sigma}n_{0j-\sigma}^d +
\beta_{j\sigma}m_{j-\sigma}}{1+\beta_{\sigma}}\label{eq14.2} \\
F_{j\sigma}^{(3)}&=&-\frac{\alpha_{j\sigma}n_{0j-\sigma}^d +\beta_{j\sigma}(n_{0j-\sigma}^d
-m_{j-\sigma} ) }{1-\beta_{\sigma}}
\label{eq14.3}
\end{eqnarray}
where $n_{0j\sigma}^d$, $m_{j\sigma}$,
$\alpha_{j\sigma}$ and $\beta_{j\sigma}$ are given in Ref. \cite{ref4}. The correlation 
function $F_{j\sigma}^{(2)}$ is high doping $\delta$ dependent (where $\delta =1-n_T$), 
mainly near half-filling. 
Considering a model where $t_{0j}^d=t^d$ for the $z$ neighbors, only one value $(j=1)$ 
of $h_{2j\sigma}$, named $h_{21\sigma}$, is need. 
\begin{figure}[!ht]
\begin{center}
\includegraphics[angle=-90]{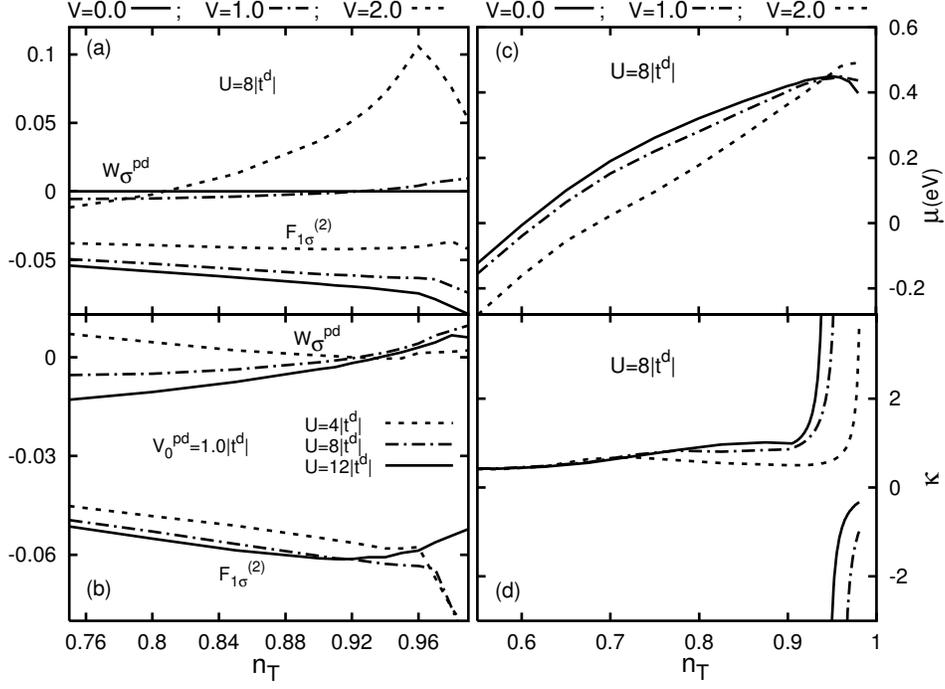}
\end{center}
\caption{In (a), the hybridized shift $W_{\sigma}^{pd}$ and the correlation function 
$F_{1\sigma}^{(2)}$ are shown as functions of $n_T$ for $U=8|t^d|$. The hybridizations
$V=V_0^{pd}/|t^d|$ are given in the top of figure (a).
(b) show $W_{\sigma}^{pd}$ and $F_{1\sigma}^{(2)}$ for $V_0^{pd}=1.0|t^d|$ and different 
Coulomb interactions. In (c), the chemical potential is shown as a function of $n_T$ for 
$U=8|t^d|$. The the hybridizations $V$ are given in the top of figure (c). 
(d) dependence of the compressibility on $n_T$ for the parameters identical to (c). 
($k_BT=0.0$ and $t^d=-0.5eV$).}
\label{fig1}
\end{figure}
The figures  \ref{fig1}(a) and \ref{fig1}(b) show the behavior of $W_{\sigma}^{pd}$ and 
$F_{1\sigma}^{(2)}$ as functions of $n_T$ for different values of Coulomb interaction 
$U$ and $V_0^{pd}$. The quantity $V_0^{pd}$ stand for the $\bf k$-independent hybridization 
\cite{ref4}. It is clear from figures \ref{fig1}(a) and \ref{fig1}(b) that at least two 
important effects presented in the band shift $W_{{\bf k}\sigma }$ are responsible for the 
unexpected behavior of the compressibility. 
The first effect is associated with the spin-correlation term $\langle S_j^+S_0^-\rangle$, which decreases 
significantly near half-filling as can be observed in figures \ref{fig1}(a) and \ref{fig1}(b). 
Therefore, due to the effect of $\langle S_j^+S_0^-\rangle$, in the underdoping regime, the lower 
Hubbard band becomes very flat in the region of the point $(\pi,\pi)$ \cite{ref52}, resulting 
in a peak in the density of states close to the charge-transfer gap (due to Coulomb 
interaction $U$). As consequence, below a critical doping $\delta_c$, the chemical potential 
$\mu$ decreases. Moreover, the compressibility diverges in $\delta_c$, and then, it becomes negative
(see figs. \ref{fig1}(c) and \ref{fig1}(d) ). Nevertheless, it has been verified that the hybridized shift 
$W_{\sigma}^{pd}$ presented in $W_{{\bf k}\sigma}$ 
moves the divergence towards the half-filling  when the hybridization 
is enhanced. This behavior is clear in figures \ref{fig1}(c) and \ref{fig1}(d), which show the 
chemical potential and the compressibility as functions of $n_T$.
Using a systematic large-$N$ expansion, in Ref. \cite{ref6} the authors have studied 
the compressibility of the two-dimensional infinite-$U$ one-band Hubbard model. They showed 
that a critical doping $\delta_c=0.07\pm 0.01$ exists, and it may sign the instability of the 
Fermi liquid phase (below $\delta_c$) and the onset where phase separation can take place. 
Concerning the divergence in the compressibility, although the present model and procedure  
are distinct from those used in reference \cite{ref6}, the results are in agreement. 
Finally, it has been also verified 
that the $d-p$ hybridization moves the divergence in the compressibility towards the
half-filling, mainly due to its effect on $W_{\sigma}^{pd}$. Therefore, the doping range 
where there are phase separation and Fermi liquid instability  decreases 
if the hybridization is enhanced.


\newpage
\begin{thebibliography}{99}
\bibitem{ref00}
J. L. Tallon and J. W. Loram, Physica C, {\bf 349} (2001) 53.

\bibitem{ref01}
E. Dagotto, Rev. Mod. Phys., {\bf 66} (1993) 763.

\bibitem{ref4}
E.J. Calegari, S.G. Magalh\~aes and A.A.\ Gomes, Eur. Phys. J. B  {\bf 45} (2005) 485. 

\bibitem{ref5}
L.M. Roth, Phys. Rev. {\bf 184} (1969) 451.

\bibitem{ref2}
E.J. Calegari, S.G. Magalh\~aes and A.A.\ Gomes, Intern.
Journ. of Modern Phys. B, {\bf 18} (2004) 241.

\bibitem{ref3}
E.J. Calegari, S.G. Magalh\~aes and A.A.\ Gomes, Physica B {\bf 359 - 361C} (2005) 560.

\bibitem{ref52}
J. Beenen and D.M. Edwards, Phys. Rev. B {\bf 52} (1995) 13636.

\bibitem{ref6}
Arti Tandon {\it et al.}, Phys. Rev. Lett. {\bf 83} (1999) 2046.


\end{thebibliography}
\end{document}